\documentclass[apl,reprint,twocolumn,floatfix]{revtex4}
\usepackage{graphicx}
\usepackage{bm}
\usepackage{textcomp}
\usepackage{color}

\begin{document}

\title{Low-loss tunable metamaterials using superconducting \\circuits with Josephson junctions}

\author{P.~Jung}
\affiliation{Physikalisches Institut, Karlsruhe Institute of Technology,
D-76128 Karlsruhe, Germany}

\author{S.~Butz}
\affiliation{Physikalisches Institut, Karlsruhe Institute of Technology,
D-76128 Karlsruhe, Germany}

\author{S.~V.~Shitov}
\affiliation{National University of Science and Technology MISIS, Moscow 119049, Russia}
\affiliation{Institute of Radio Engineering and Electronics (IREE RAS), Moscow 125009, Russia}

\author{A.~V.~Ustinov}
\email{ustinov@kit.edu}
\affiliation{Physikalisches Institut, Karlsruhe Institute of Technology,
D-76128 Karlsruhe, Germany}
\affiliation{National University of Science and Technology MISIS, Moscow 119049, Russia}

\date{\today}

\begin{abstract}
We report on experiments with superconducting metamaterials containing Josephson junctions. In these structures, split-ring resonators used in conventional metamaterials are replaced by superconducting loops that are interrupted by Josephson junctions, so called rf-SQUIDs. Like the split-ring resonators, these elements can be seen as LC-resonators that couple to the magnetic field. The advantage of superconducting thin-film metamaterials is that, due to the tunable intrinsic inductance of the Josephson junction, the resonance frequency of the rf-SQUID can be changed by applying an external dc magnetic field. We present experimental results that demonstrate the tunability of the resonance frequency of these devices.
\end{abstract}



\maketitle
\noindent
One of the main limitations of many metamaterial designs is the restriction of their usability to a narrow frequency band. By using resonant elements in order to achieve the desired effects (such as a negative magnetic permeability $\mu_r$) the “meta-atoms” are usually made for a fixed operation frequency range by design. To circumvent this limitation, several approaches have been developed \cite{boardman2011}, many of which are based on the idea of introducing a nonlinear element into the resonant circuit, thus making it tunable \cite{gil2004,kapitanova2011}. Superconductors in general and Josephson junctions in particular are ideal constituents of tunable meta-atoms due to the strong dependence of their inductance on external parameters such as temperature and magnetic field. The tunability by varying temperature has already been demonstrated with superconducting split-ring resonators (SRRs)\cite{ricci2007,wu2011}. In this work, we move one step further and experimentally demonstrate tunability via magnetic field by replacing the SRR by a superconducting quantum interference device (SQUID). This approach has previously been suggested in the theoretical work by Lazarides and Tsironis \cite{lazarides2007} and further investigated by Maimistov and Gabitov   \cite{gabitov2012}. 

\begin{figure}[b]
\begin{center}
\includegraphics[width=0.8\linewidth]{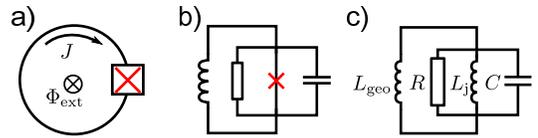}
\end{center}
\caption{(a) A superconducting loop interrupted by a Josephson junction can be decomposed into a (b) network model using the restively and capacitively shunted junction model (RCSJ) \cite{tinkham}. The red cross indicates the Josephson junction for which voltage and current relation are given by the Josephson relations. (c) Small signal network model of the SQUID. $L_{\rm geo}$ presents the geometric inductance of the superconducting loop. $L_{\rm j}$ is the Josephson inductance, see Eq.\,(\ref{jind}). $R$ and $C$ are the normal resistance and capacitance of the Josephson junction, respectively. In our sample, the junction is shunted by a large additional capacitance to decrease the resonance frequency, which is added in parallel to $C$.}
\label{fig:squid_lc_model}
\end{figure}

The key element used in this work is the Josephson junction which consists of two superconductors separated by a thin insulating layer. The current and voltage over such a junction is described by the two Josephson relations \cite{josephson1962,likharev,tinkham}. The Josephson junction with a critical current $I_{\rm c}$ can be treated as a nonlinear network element. A simple description is provided by a small signal approximation. A small variation $\delta \varphi$ of the phase difference $\varphi$ between the superconducting wave functions on both sides of the junction around a constant value $\varphi'$ is translated into a small variation of the current $I = I_{\rm c} \sin \varphi' + \delta I$ flowing through it. In this way, current and voltage are related to each other by \cite{likharev,tinkham}:
\begin{equation}\label{jind}
 V  = \frac{\Phi_0}{2 \pi I_{\rm c} \cos \varphi' } \frac{d}{dt} I  \equiv L_{\rm j} (\varphi') \frac{d}{dt}I
\end{equation}

This equation takes the form characteristic to an inductive element. That is, the voltage over the element is proportional to the time derivative of the current through it. In this model, the value of the so-called Josephson inductance $L_{\rm j}$ depends on the constant part of the phase difference across the junction and therefore on the bias current through it. Thus, it works as a tunable inductor. 

A circuit consisting of a Josephson junction placed into a superconducting loop (cf. Fig.\,\ref{fig:squid_lc_model}a) is  a single junction (rf-)SQUID. In this case, magnetic flux quantization leads to the relation
\begin{equation}\label{fluxquantization}
\varphi \frac{\Phi_0 }{ 2 \pi }  +  L_{\rm geo} I = \Phi_{\rm ext}.
\end{equation}

The electric circuit equivalent of an rf-SQUID is depicted in Fig.\,\ref{fig:squid_lc_model}b. In addition to the Josephson relations, one has to account for the loop inductance, the capacitance between the superconducting electrodes and the (resistive) single electron tunneling current.

Once an external magnetic flux  $\Phi_{\rm ext}(t) = \Phi_{\rm e0} + \Phi_{\rm ea} \cos (\omega t)$ is applied to the system and the driving ac amplitude  $\Phi_{\rm ea} \ll I_c / L_{\rm geo}$ is small, the small signal approximation is valid. The total inductance  $L_{\rm tot}$ comprises the geometric inductance $L_{\rm geo}$ and the tunable Josephson inductance $L_{\rm j}(\Phi_{\rm e0})$, see Fig.\,\ref{fig:squid_lc_model}c. The resonance frequency of the system is given as a function of the external bias flux $\Phi_{\rm e0}$
\begin{eqnarray}
\label{LC-resonance}
\omega_0(\Phi_{\rm e0}) &=& \left ( L_{\rm tot}(\Phi_{\rm e0}) C\right ) ^ {-\frac{1}{2}}. 
\end{eqnarray}

In the present experiment, the additional kinetic inductance \cite{likharev} of the superconducting ring is negligible compared to its geometric inductance. It should be noted, though, that close to the critical temperature of the superconductor, the kinetic inductance will increase significantly and influence the resonance frequency. Due to this fact and the temperature dependence of the critical current of the junction, the rf-SQUID also shows an intrinsic tunability by temperature which, however, will not be detailed here.

\begin{figure}[b]
\begin{center}
\includegraphics[width=0.9\linewidth]{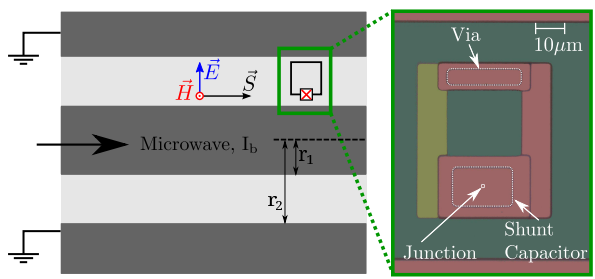}
\end{center}
\caption{Sketch of a single rf-SQUID embedded into a coplanar transmission line. The darker areas represent the Nb electrodes. The inset on the right shows an optical microscope image of a single rf-SQUID. }
\label{fig:squid_waveguide}
\end{figure}

The approximation of the Josephson junction by a tunable inductor works well within the mentioned small signal limit. However, in order to obtain a complete description, the Josephson nonlinearity and loss terms have to be included. By applying Kirchhoff's laws, the two Josephson relations and Eq.\,(\ref{fluxquantization}) to the circuit depicted in Fig.\,\ref{fig:squid_lc_model}b, a second order nonlinear differential equation is derived:

\begin{equation}\label{DGL}
 \Phi + L_{\rm geo} \left [ I_{\rm c} \sin \left ( \frac{2 \pi}{\Phi_0} \Phi \right )   +  \frac{1}{R} \dot \Phi  +  C \ddot \Phi \right ] = \Phi_{\rm ext}(\omega,t).
\end{equation}

\noindent The time average of the solution of this equation yields the complex ac flux susceptibility of the single SQUID which is $\chi_\phi = \left < \Phi / \Phi_{\rm ext} \right >_t - 1$. There is a strong frequency dependence of real and imaginary components of $\chi_\phi$ around the resonance frequency $\omega_0$. 

For an effective medium consisting of these elements, the magnetic susceptibility is given by $\chi_{\rm mag} = \tilde F \chi_\phi$. The coefficient $\tilde F$ depends on the arrangement of the meta-atoms in the medium. Provided that the coupling between the SQUIDs is small (i.e. $\tilde F \ll 1 $) \cite{lazarides2007}, the relative permeability of the medium is 
\begin{equation}\label{mu_r}
\mu_{\rm r} = 1 + \chi_{\rm mag} = 1 + \tilde F \chi_\phi.
\end{equation}

In this work, we experimentally verify the microwave properties of a prototype circuit representing a single superconducting meta-atom of such a medium. We have designed and fabricated a circuit consisting of a single rf-SQUID placed inside the gap of a coplanar waveguide (CPW). In this setting, the SQUID loop is perpendicular to the magnetic field, as can be seen in Fig.\,\ref{fig:squid_waveguide}a, and thus couples to the quasi-TEM mode of the CPW. The arrows denoted by $\vec E$, $\vec H$ and $\vec S$ indicate the orientation of the electric and magnetic field components and the Poynting vector of the incoming wave, respectively. An external magnetic field is applied perpendicular to the substrate plane either through an external coil or by driving a dc current through the central strip of the transmission line. In either case, the SQUID loop is biased with a dc magnetic flux $\Phi_{\rm e0}$ thus making the resonance frequency tunable in situ according to Eq.\,(\ref{LC-resonance}).  

\begin{figure}[t]
\begin{center}
\includegraphics[width=0.9\linewidth]{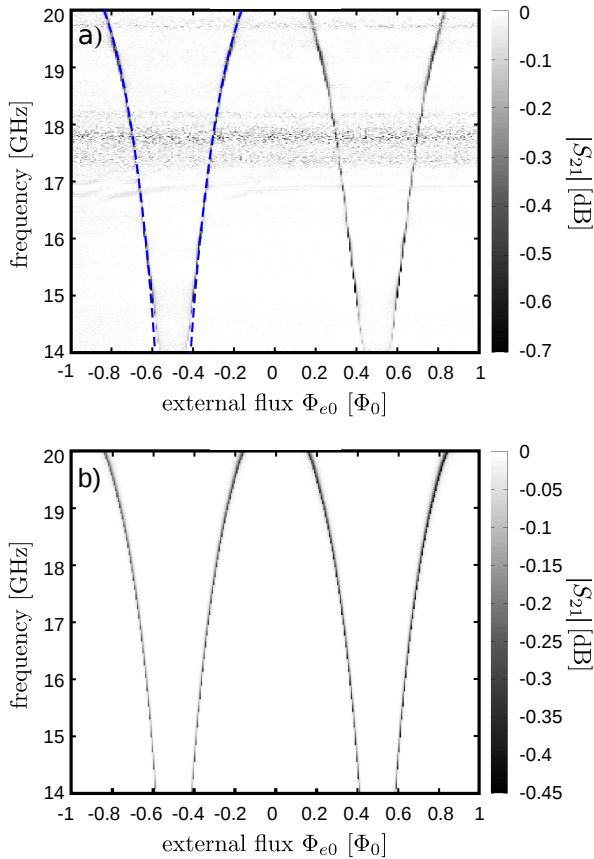}
\end{center}
\caption{(a) Measured microwave transmission at a power of $P \approx -85$\,dBm at the sample, that corresponds to $\Phi_{\rm ea} \approx \Phi_0 /1600$. The darker colors correspond to a reduced transmission through the SQUID-loaded waveguide. The blue dashed line represents the analytical solution extracted from the small signal model (Fig.\,\ref{fig:squid_lc_model}c). It is shown for negative flux values only in order not to obscure the data. (b) Transmission through a single SQUID simulated numerically by solving Eq.\,(\ref{DGL}).}
\label{fig:meas_vs_sim}
\end{figure}

The SQUID loop used in the presented measurement is made from niobium (which has a superconducting transition temperature of $T_c = 9.2$\,K) and includes a $\rm Nb$/$\rm AlO_x$/$\rm Nb$  circular Josephson tunnel junction with a radius of 0.5nm and a critical current $I_{\rm c} \approx 4.5\,\mu A$ . The geometric loop inductance $L_{\rm geo} = 63.5$\,pH is comparable to the zero-flux Josephson inductance $L_{\rm j}(0) = 74$\,pH. By shunting the Josephson junction with an additional large capacitor $C_{\rm shunt} = 1.7$\,pF we designed the resonance frequency to be tunable between about 21\,GHz and 5\,GHz, for an infinitesimally small $\Phi_{\rm ea}$. The measurements were performed using a cryogenic amplifier and a vector network analyzer while the sample was immersed in liquid $^4$He at a temperature of 4.2\,K. 

\begin{figure}[b]
\begin{center}
\includegraphics[width=0.8\linewidth]{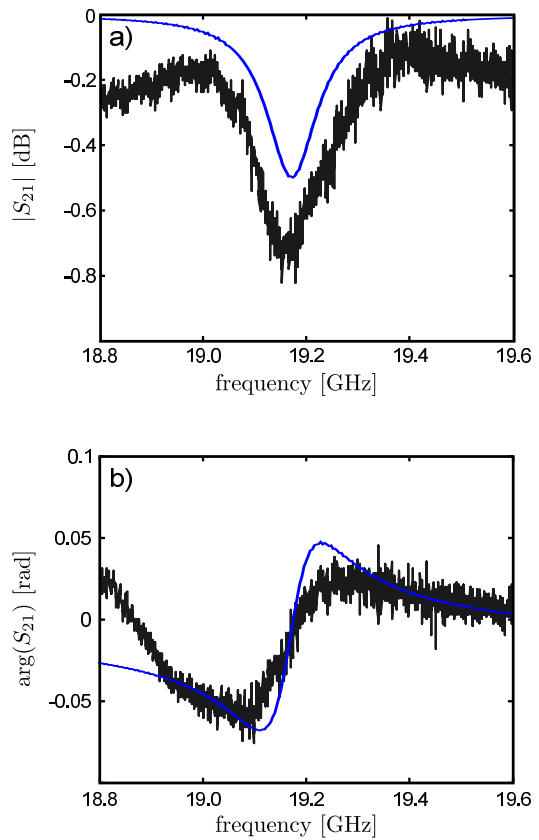}
\end{center}
\caption{Relative (a) magnitude and (b) phase of the transmission ($S_{\rm 21}$) through the SQUID-loaded coplanar waveguide at a bias of $\Phi_{\rm e0} = 0.77\, \Phi_0$. The measured data (power $P \approx  -110$\,dBm) are shown in black, the simulated curve ($\Phi_{\rm ea} = \Phi_0 / 2^{15}$) in blue.}
\label{fig:single_trace}
\end{figure}

The experimental results for low-power transmission presented in Fig.\,\ref{fig:meas_vs_sim}a demonstrate the frequency tunability of the resonance dip. The decreased transmission is illustrated by the darker color in the plot. The resonance frequency changes periodically as a function of the applied magnetic flux $\Phi_{\rm e0}$ with a period of $\Phi_{\rm 0}$. The dashed blue line represents a fit of the small signal model of Eq.\,(\ref{LC-resonance}). As discussed earlier, the model should yield good fit results only for very small values of the driving amplitude. The power used in this measurement was -85\,dBm which corresponds to an ac flux amplitude of $\Phi_{\rm ea} \approx \Phi_0 / 1600$ and can be safely assumed to be small enough for the approximation to hold. The values, obtained from the fit for $L_{\rm geo}$ and $C$ agree within 5\% with the values expected for the fabrication procedure while the value of the critical current $I_{\rm c}$ is about 6\% less than the one measured on test junctions on the same chip. 

In order to estimate the magnetic susceptibility and thereby the relative permeability, Eq.\,(\ref{DGL}) was solved numerically. The parameters $L_{\rm geo}$ and $C$ were taken from the fit mentioned in the paragraph above. The value for the shunting resistance $R$ was taken from the current-voltage characteristics measured for a test junction on the same chip. Additionally, the loss tangent of the $\rm Nb_2O_5$ dielectric of the shunt capacitor was assumed to be $\tan \delta = 1.0 \cdot 10^{-3}$ \cite{kaiserphd}. To calculate the scattering characteristics versus dc magnetic flux $\Phi_{\rm e0}$, we employed the effective medium approach by calculating the effective $\mu_{\rm r}$ of a SQUID-loaded section of the CPW. Analogous to the procedure in Ref.~\cite{lazarides2007}, our task was to find a suitable value for $\tilde F$. Since the actual field distribution in a superconducting coplanar waveguide does not have a simple, analytical description, we approximated the system by treating it as a homogeneous, cylindrically symmetric problem, similar to a coaxial cable. The unit cell of the waveguide, of length $l$, contains one SQUID which defines the interaction with the magnetic field component. Under these assumptions 
\begin{equation}\label{our_ftilde}
 \tilde F = \frac{\mu_0 A_{\rm sq}^2 }{L_{\rm geo} \pi l  \left ( r_2^2 - r_1^2\right ) }.
\end{equation}
Here, $A_{\rm sq}$ is the area of the SQUID, $r_2$ and $r_1$ are the outer and inner radius of the gap, as shown in Fig.\,\ref{fig:squid_waveguide}a. 

Using Eqs.\,(\ref{mu_r}) and (\ref{our_ftilde}), we obtain the effective permeability $\mu_{\rm r}$ and use it to modify the inductance and thus the impedance of the CPW section. This, in turn, allows us to calculate the scattering matrix coefficient $S_{21}$. The result of this calculation, depicted in Fig.\,\ref{fig:meas_vs_sim}b, is in excellent agreement with the measured data. 
As expected, at so small applied power level, the numerical treatment adds very little additional information to the analytical solution of Eq.\,(\ref{LC-resonance}) shown in Fig.\,\ref{fig:meas_vs_sim}a by the dashed blue line. However, at higher power levels, the small signal approximation of Eq.\,(\ref{jind}) is no longer valid and the numerical treatment will be relevant.

The frequency-dependent properties of a metamaterial can be characterized by relative magnitude and phase of the complex transmission coefficient $S_{\rm 21}$.  Measurement and numerical calculation of the transmission magnitude and phase through the SQUID-loaded coplanar waveguide (cf. Fig.\,\ref{fig:single_trace}a and \ref{fig:single_trace}b) are in good qualitative agreement. Detailed quantitative comparison in our case is hindered by difficulty of making precise calibration measurements at cryogenic temperatures.

In conclusion, we have experimentally and theoretically investigated a prototype of a magnetically coupled, tunable meta-atom made of a thin-film superconducting circuit. The resonance frequency of a superconducting loop containing a Josephson junction is tunable over a wide frequency range both by magnetic field and temperature. Therefore, the SQUID metamaterial appears as an appealing alternative to the widespread normal-metal structures employing split-ring resonators. Further studies of SQUID properties at higher powers, i.e. investigating its nonlinear properties, as well as experiments on 1D and 2D-arrays of such elements are presently being conducted.

The authors would like to acknowledge stimulating discussions with S. M. Anlage, I. Gabitov, and G. Tsironis.  This work was supported in part by the Ministry of Education and Science of the Russian Federation, and also by the Deutsche Forschungsgemeinschaft (DFG) and the State of Baden-W{\"u}rttemberg through the DFG Center for Functional Nanostructures (CFN).  Philipp Jung would like to acknowledge the financial support by the Helmholtz International Research School for Teratronics (HIRST), Susanne Butz would like to acknowledge the financial support by the Landesgraduiertenförderung Baden-W\"urttemberg.

\end{document}